%% file: PPMAvaccine.tex
\documentclass[12pt]{article}
\usepackage[top=1in, bottom=1in, left=1in, right=1in]{geometry}
\usepackage{graphicx}
\usepackage{setspace}
\usepackage{amssymb,amsmath}
\usepackage{booktabs}
\usepackage{multirow}
\usepackage[round,authoryear]{natbib}
\usepackage{lscape}
\usepackage{url}
\usepackage{fancyhdr}

\bibpunct{(}{)}{;}{a}{}{,}

\title{Using Proxy Pattern-Mixture Models to Explain Bias \\in Estimates of COVID-19 Vaccine Uptake \\from Two Large Surveys}
\author{Rebecca R. Andridge \\The Ohio State University College of Public Health\\1841 Neil Ave., Columbus, OH 43220\\andridge.1@osu.edu}
\date{July 31, 2023}

\begin{document}

\maketitle


\begin{abstract}
Recently, attention was drawn to the failure of two very large internet-based probability surveys to correctly estimate COVID-19 vaccine uptake in the United States in early 2021. Both the Delphi-Facebook CTIS and Census Household Pulse Survey (HPS) overestimated uptake substantially, by 17 and 14 percentage points in May 2021, respectively. These surveys had large numbers of respondents but very low response rates ($<$10\%), thus, non-ignorable nonresponse could have had substantial impact. Specifically, it is plausible that  ``anti-vaccine" individuals were less likely to participate given the topic (impact of the pandemic on daily life). In this paper we use \emph{proxy pattern-mixture models (PPMMs)} to estimate the proportion of adults (18+) who received at least one dose of a COVID-19 vaccine, using data from the CTIS and HPS, under a non-ignorable nonresponse assumption. Data from the American Community Survey provide the necessary population data for the PPMMs. We compare these estimates to the true benchmark uptake numbers and show that the PPMM could have detected the direction of the bias and provide meaningful bias bounds. We also use the PPMM to estimate vaccine hesitancy, a measure for which we do not have a benchmark truth, and compare to the direct survey estimates.
~\\
\textbf{Running Head: }\emph{Using PPMMs to Estimate COVID-19 Vaccine Uptake}
\end{abstract}

\doublespace


\section{Introduction}

In the absence of nonresponse, carefully designed probability samples provide a principled way of producing unbiased estimates of population quantities such as proportions and means. Random selection of individuals into a sample, where every population unit has a known, non-zero probability of selection, ensures that the sample represents the population in expectation. Federal statistical agencies in the United States and abroad rely on such surveys to produce official estimates of population-level characteristics that play an important in policy-making and business strategies \citep{Hastak2001}. These large, government-sponsored surveys are generally large and expensive, requiring years of development (e.g., field-testing) as well as careful post-survey analysis before official statistics are released.

The COVID-19 pandemic created a unique challenge in that it created a sudden, unanticipated need for data to describe both the incidence of disease and how the pandemic was impacting daily life. In this paper we analyze two large surveys that were implemented quickly in response to the pandemic: the U.S. Census Bureau's Household Pulse Survey (HPS) \citep{Fields2020} and the Delphi-Facebook COVID-19 Trends and Impact Survey (CTIS) \citep{Salomon2021}. The HPS was a government-sponsored survey, whereas the CTIS was a collaboration between academia and a private company. Both surveys were large probability samples that repeatedly collected information on a range of pandemic-related topics; we focus on the estimation of vaccine uptake in early 2021 when vaccines first became available in the U.S.. Average sample sizes (number of respondents) for the HPS was approximately 75,000 per wave and for CTIS it was approximately 250,000 per week.

\begin{figure}[tb]
\centering
\includegraphics[width=0.8\textwidth]{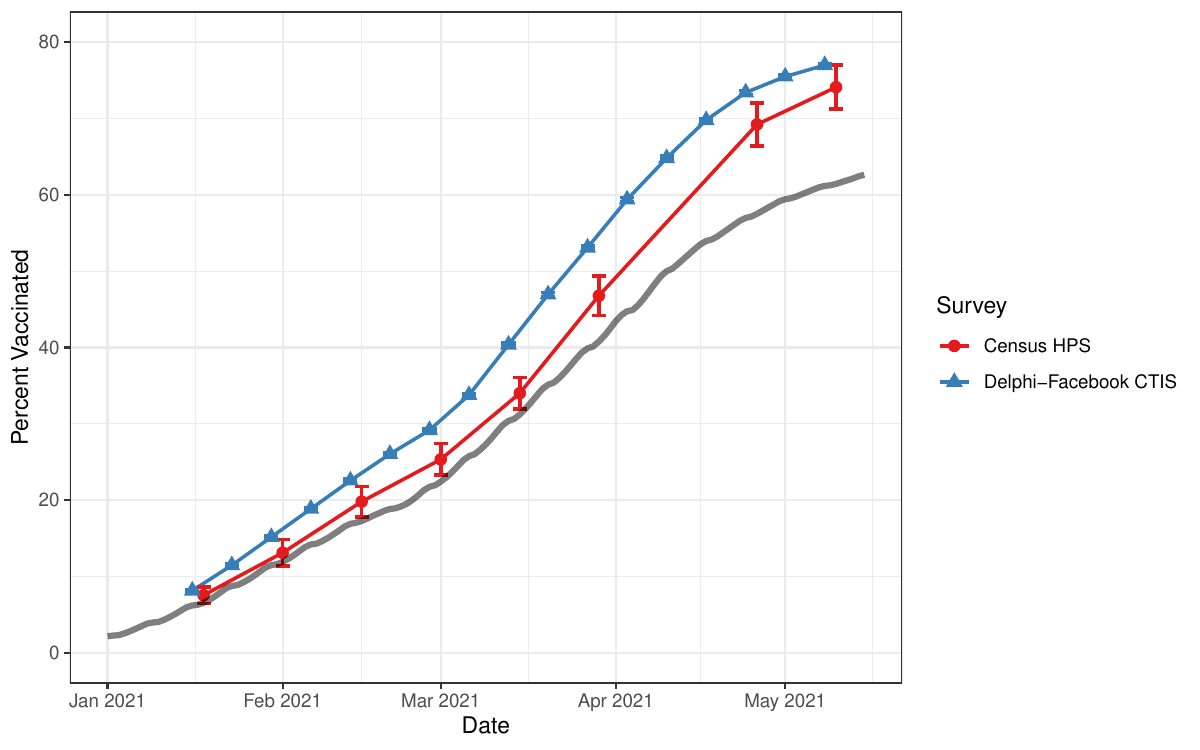}
\caption{Survey weighted estimates of COVID-19 vaccine uptake for adults in the U.S. in 2021 compared to CDC benchmark data (grey line), plotted by the end date of each survey wave. Intervals are 95\% CIs; for Delphi-Facebook CTIS the CIs are too small to be visible.}
\label{fig:misses}
\end{figure}

Despite their large sizes, both the Census HPS and Delphi-Facebook CTIS produced substantially biased estimates of vaccine uptake in the U.S. in early 2021 \citep{Nguyen2021,Bradley2021}. As shown in Figure~\ref{fig:misses}, the weighted estimates from these surveys consistently overestimated vaccine uptake (the percentage of U.S. adults reporting receiving at least one dose of a COVID-19 vaccine) as compared to benchmark data retrospectively available from the U.S. Centers for Disease Control and Prevention (CDC) \citep{CDCdata}. \citet{Bradley2021} decomposed the error in the survey estimates of vaccine update for both surveys using the framework of \citet{Meng2018}, emphasizing the danger of very large samples leading to very precise (negligible confidence interval length) but severely biased results.

Importantly, while these two surveys resulted in large samples, they had very small response rates. In the period from January through May 2021, unweighted response rates for the HPS were in the range of 6.6-7.8\% \citep{HPStechdoc}. Response rates are not available for the CTIS, but daily cooperation rates\footnote{Percent of all Facebook users who logged onto Facebook and received an invitation on a given day who then answered the symptoms question on that day} were approximately 0.5-1.5\% \citep{CITSmethods}. With such small response rates, the protection against bias afforded by probability sampling is erased, and these surveys in many ways resemble nonprobability samples (e.g., convenience samples). A detailed analysis of nonresponse for the HPS \citep{HPSNR2020} showed that response rates differed across demographic domains (e.g., age, race, ethnicity). Post-survey weighting adjustments were used for both surveys to attempt to correct for differential nonresponse, but were limited to a small set of demographic characteristics. Given that these weighting adjustments failed to produce unbiased estimates, and with such small response rates, we hypothesized that a \emph{non-ignorable} nonresponse mechanism might have been responsible at least in part for the biased estimates.

In the context of measuring vaccine uptake, if an individual's propensity to respond to either the HPS or CTIS is at least in part a function of their vaccine status, this constitutes a \emph{non-ignorable} nonresponse mechanism. Specifically, it is plausible that people who were ``anti-vaccine" (and thus were unvaccinated) were less likely to complete these surveys on the impact of the COVID-19 pandemic on daily life. One could also hypothesize that individuals who were anti-vaccine might also be suspicious of the government and thus less likely to respond to the HPS, which was an official government-sponsored survey.

In order to assess whether this type of non-ignorable nonresponse may have been occurring, we use previously developed \emph{proxy pattern-mixture models (PPMMs)} \citep{AndridgeLittle2011,AndridgeLittle2020}, which allow for estimation under a non-ignorable nonresponse assumption, to estimate vaccine uptake using data from both surveys. In Section~\ref{sect:surveys} we describe the HPS and CTIS in more detail. In Section~\ref{sect:ppma} we briefly review the PPMM, and present results from applying it to estimate vaccine uptake in Section~\ref{sect:uptake}. In Section~\ref{sect:hesitant} we use the PPMM to estimate vaccine hesitancy, a measure for which we do not have a benchmark truth. We conclude in Section~\ref{sect:discussion} with discussion of how the PPMM could have been used prospectively as part of a nonresponse bias assessment and describe factors that would facilitate such analyses in the future.

\section{Details on the COVID-19 Vaccine Surveys}\label{sect:surveys}

\subsection{Census Household Pulse Survey}

The Census Household Pulse Survey (HPS) was an experimental data product of the U.S. Census Bureau that was developed in the early phase of the COVID-19 pandemic in conjunction with (\url{https://www.census.gov/data/experimental-data-products/household-pulse-survey.html}). The first phase of this survey launched on April 23, 2020 with the goal of quickly and efficiently collecting data about how the pandemic was affecting the lives of individuals residing in the United States, and was still ongoing as of March 2023. Survey questions asked about experiences that may be affected by the pandemic, with a focus on employment status, food security, housing security, physical and mental health, and educational disruption \citep{Fields2020}. Starting in January 2021, when COVID-19 vaccines became available, questions were added about vaccination status and intention. Table~\ref{tab:questions} lists the questions used to estimate vaccine uptake and vaccine hesitancy.

Given the goal of quick survey deployment and results dissemination as well as the context (during the pandemic), all data collection was via web. The HPS consisted of repeated, stratified, cross-sectional random samples with a target population of all adults (18+) residing in housing units in the U.S. (excluding Puerto Rico). As with many demographic surveys conducted by federal statistical agencies, the HPS sampled households from the Census Bureau's Master Address File (MAF). However, due to the online-only design, only addresses on the MAF that had a linked cell phone number and/or email address (from the Census Bureau Contact Frame) were eligible for sampling due to the online-only survey design. Approximately 80\% of housing units on the MAF had a cell phone and/or email address (\url{https://www.census.gov/programs-surveys/household-pulse-survey/technical-documentation.html}). Initially, samples were drawn weekly from the MAF, with a shift to bi-weekly samples in August 2020. The sample was stratified by geographic area (50 states, Washington D.C., top 15 Metropolitan Statistical Areas). Sampled individuals were contacted by text and/or email with a request to complete the survey.

\begin{table}[tb]
\begin{center}
\caption{Survey questions about vaccine uptake and intention in the Census Household Pulse Survey and Delphi-Facebook CTIS, January 2021 - May 2021}
\label{tab:questions}\begin{tabular}{lp{0.9\textwidth}}
\toprule
\multicolumn{2}{l}{\textbf{Census Household Pulse Survey}}\\ \midrule
Uptake & Question: ``Have you received a COVID-19 vaccine?" \\ \cmidrule{2-2}
& Response Options: ``Yes", ``No"\\ \midrule
Intention & Question: ``Once a vaccine to prevent COVID-19 is available to you, would you\dots" [only asked if did not respond ``Yes" to uptake question]\\  \cmidrule{2-2}
&  Response Options:  ``Definitely get a vaccine", ``Probably get a vaccine", ``Be unsure about getting a vaccine"*, ``Probably NOT get a vaccine", ``Definitely NOT get a vaccine"\\ \midrule
\multicolumn{2}{l}{\textbf{Delphi-Facebook CTIS}}\\ \midrule
Uptake & Question: ``Have you had a COVID-19 vaccination?" \\  \cmidrule{2-2}
& Response Options: ``Yes", ``No", ``I don’t know" \\ \midrule
Intention & Question: ``If a vaccine to prevent COVID-19 were offered to you today, would you choose to get vaccinated?" [only asked if did not respond ``Yes" to uptake question]\\  \cmidrule{2-2}
& Response Options:  ``Yes, definitely", ``Yes, probably", ``No, probably not", ``No, definitely not"\\ \midrule
\multicolumn{2}{l}{\small{*option added mid-April 2020}}
\end{tabular}
\end{center}
\end{table}

We analyzed iterations of the HPS conducted from January 6, 2021 through May 10, 2021. During this time period, approximately 1,000,000 housing units were sampled in each data collection period with 68,000-80,000 respondents per wave.

Several post-survey adjustments were made to the HPS base weights to produce the final analytic weights, including adjustments for nonresponse, undercoverage, and a conversion from household-level to person-level weights \citep{Fields2020}. As a last step, an iterative raking procedure was used to ensure that weighted totals match the U.S. adult population with respect to specified demographic characteristics. Specifically, weights were raked to two sets of population totals from the 2019 American Community Survey: educational attainment by age and sex\footnote{Surveys conducted by the U.S. federal government historically have collected sex as a binary variable and without nuance, i.e., conflating it with gender. We acknowledge this limitation.} within state, and race/ethnicity by age and sex within state.

\subsection{Delphi-Facebook COVID-19 Trends and Impact Survey}

The Delphi-Facebook COVID-19 Trends and Impact Survey (CTIS) (\url{https://delphi.cmu.edu/covid19/ctis/}) was developed in the early phase of the COVID-19 pandemic as a collaboration between Meta (Facebook's parent company) and the University of Maryland and Carnegie Mellon University \citep{Barkay2020}. The survey launched on April 6, 2020 and ended on June 25, 2022. The stated main goal of the surveys was to collect real-time indicators of symptom severity, both individual and household-level \citep{Kreuter2020}. Starting in January 2021, questions about vaccination status and intention were added, with the exact wording as shown in Table~\ref{tab:questions}.

The CTIS was a large, stratified, cross-sectional random samples, drawn daily, with a target population of all adults (18+). The survey was implemented in over 200 countries; we only use data from the U.S. in our analyses. The sampling frame was all Facebook users (18+) who had been active on Facebook in the previous month. Samples were drawn daily, and the survey invitation was shown at the top of the Facebook feed for selected individuals\citep{Salomon2021}. In the U.S. the sample was stratified by state.

We pooled the daily CTIS samples into weeks and analyzed the weeks ending in January 16, 2021 through May 8, 2021. During this time period, an average of approximately 290,000 respondents provided at least partial responses to the survey each week.

Multiple post-survey adjustments were made to the CTIS base weights to account for nonresponse and non-coverage (due to the fact that not all of the target population are Facebook users) \citep{CITSweights}. First, inverse propensity score weighting was used to adjust for nonresponse within the sampling frame (the Facebook user base) using age and gender as predictors of response status. Then post-stratification was used to ensure weighted totals match the target population with respect to age by sex within state using the Current Population Survey 2018 March Supplement for population totals \citep{Barkay2020}.

\section{Methodology: The Proxy Pattern-Mixture Model} \label{sect:ppma}

The proxy pattern-mixture model (PPMM) was originally proposed by \citet{AndridgeLittle2011} as a tool for assessing the potential impact of non-ignorable nonresponse on estimating means of continuous variables, primarily in the context of item nonresponse. It was subsequently extended to estimating proportions by \citet{AndridgeLittle2020}. The PPMM has also been used as the basis for indices that quantify the potential for non-ignorable selection bias for means \citep{LittleWest2020}, proportions \citep{Andridge2019}, and regression coefficients \citep{West2021} estimated from nonprobability samples. Our goal is to estimate a proportion -- the proportion of U.S. adults who have had at least one dose of a COVID-19 vaccine -- thus we use the binary PPM of \citet{AndridgeLittle2020} in our analyses. We briefly describe their methodology here in the context of estimating vaccine uptake and refer readers to \citet{AndridgeLittle2020} for additional details.

Let $Y_i$ be the binary indicator of whether individual $i$ in the population of U.S. adults (18+) has received at least one dose of a COVID-19 vaccine. A single iteration of either the HPS or CTIS collects $Y_i$ from a subset of the population, and let $S_i$ be the sample inclusion indicator that takes the value $S_i=1$ if the individual is sampled and responds (provides a value of $Y_i$) and 0 otherwise. Since only a small fraction of sampled individuals responded to the survey, the $S_i$ we observe is a combination of the design-based sample inclusion probability (which we know) and an unknown response propensity (which we do not know). Thus the probability density of $S_i$ is unknown without additional assumptions. In the PPMM analysis we will make assumptions about the distribution of $S_i$ through a principled sensitivity analysis. In what follows, we refer to the units with $S=1$ as the ``responding sample," and note that the units with $S=0$ include both individuals who were sampled but did not respond and individuals who were not sampled.

Crucial to the implementation of the PPMM, we must also observe covariate information at the individual level for the responding individuals and in aggregate for the rest of the population. Let $Z_i=(Z_{i1}, Z_{i2}, \dots Z_{ip})$ be a set of $p$ covariates collected on the survey, which for our purposes will be limited to information we can also obtain in aggregate for the U.S. population, i.e., demographic data. In the PPMM approach, this covariate data for respondents is reduced to a single \emph{proxy} variable $X$ by regressing $Y$ on $Z$ using a probit regression model and taking $X$ to be the estimated linear predictor from this regression. Importantly, individual-level values of $X_i$ are available for all responding individuals $(S_i=1)$, as their $Z$ values can be plugged into the estimated probit regression equation. We do not observe $X_i$ for nonresponding individuals, but if we have the mean and variance of $Z$ for this part of the population from an external source then we can estimate the mean and variance of $X$ for the nonresponding portion of the population. Despite the large sample sizes of the HPS and CITS surveys, the samples are considerably smaller than the size of the full population, i.e., sampling fractions are small. Therefore, estimates of the mean and variance of $Z$ for the entire population of U.S. adults are effectively the same as estimates for the part of this population that did not respond to a single wave or week of the HPS or CTIS surveys.

The basic idea of the PPMM is that we can measure the degree of bias present for the respondent sample mean of the proxy $X$ by comparing it to the population-level mean of $X$ (based on the aggregate information for $Z$). If $X$ is correlated with $Y$, then this provides some information about the potential bias in the respondent sample mean of $Y$. If $X$ and $Y$ are highly correlated, then a small bias in $X$ suggests (but does not guarantee) a small bias in $Y$. If, however, $X$ and $Y$ are weakly correlated (which would occur if the covariates $Z$ that create $X$ are not very predictive of $Y$) then we simply do not have much evidence for or against bias in the respondent sample mean of $Y$. Fortunately, many studies have shown that demographics available in aggregate at the national level such as age, sex, race/ethnicity, and education are moderately associated with COVID-19 vaccine acceptance \citep[e.g.,][]{Reiter2020,Haile2022}.

The PPMM does not directly model $Y$ and $X$, but instead introduces a normally distributed latent variable, $U$, such that $Y=1$ when $U>0$, and models the joint distribution of $U$ and $X$. Specifically, \citet{AndridgeLittle2020} use a bivariate normal pattern-mixture model for the joint distribution of $U$ and $X$ given $S$, in which the mean and variance parameters are distinct for $S=1$ and $S=0$. Parameters of this joint distribution are fully identified for the responding sample, with the exception of the mean and variance of the latent $U$ which cannot be separately identified; as in \citet{AndridgeLittle2020} we fix the variance of $U$ at one. For the nonresponding portion of the population $(S=0)$ we can  identify the mean and variance of $X$, but not the parameters describing the distribution of $U$ or the correlation between $X$ and $U$.

The unidentified parameters of the PPMM can be identified by making an assumption about the distribution of $S$ and with the introduction of a sensitivity parameter, $\phi$. \citet{AndridgeLittle2020} show that the PPMM is just identified if we assume that the probability an individual is sampled and responds is an unspecified function of a known linear combination of $X$ and $U$, plus potentially other observed covariates $V$ that are independent of $U$ (and $Y$) and $X$:
\begin{equation} \label{eqn:restrict}
\Pr(S=1|U,X,V) = g\left((1-\phi)X^* + \phi U, V)\right)
\end{equation}
\noindent Here $X^*$ is the proxy, $X$, rescaled to have the same variance as $U$ for $S=1$, and $\phi \in [0,1]$ is the sensitivity parameter. For a specified value of $\phi$, the parameters of the PPMM are just identified, and thus the overall mean of $Y$ can be estimated as a weighted (by the responding fraction) average of estimates of $E[Y|S=1]=E[U>0|S=1]$ and $E[Y|S=0]=E[U>0|S=0]$. Though there is no information in the data with which to estimate $\phi$, certain values of $\phi$ correspond to specific types of response mechanisms, thus enabling a reasonable, bounded sensitivity analysis. Specifically, $\phi=0$ corresponds to a missing at random assumption \citep{Rubin1987}, where the probability of response is only a function of $X$ and $V$, which are observed -- this is an ignorable response mechanism. If $\phi>0$, then response depends at least in part on $U$, and therefore on $Y$ -- a non-ignorable response mechanism.

\citet{AndridgeLittle2020} provide an explicit formula for the overall mean of $Y$ under the PPMM as a function of the parameters of the underlying normally-distributed latent $U$ for respondents $(\mu_u^{(1)})$ and nonrespondents $(\mu_u^{(0)}, \sigma_{uu}^{(0)})$ and the fraction of the population that responded ($\pi$),
\begin{align}
\mu_y &{}= \pi \Phi \left(\mu_u^{(1)} \right) + (1-\pi) \Phi \left(\mu_u^{(0)}\big/\sqrt{\sigma_{uu}^{(0)}} \right), \label{eqn:ppmm1}
\end{align}
\noindent where $\Phi(z)$ denotes the CDF of the standard normal distribution evaluated at $z$. With the identifying restriction in \eqref{eqn:restrict}, the mean and variance of $U$ for nonrespondents are given by
\begin{align}
\mu_u^{(0)} &{}= \mu_u^{(1)} + \left( \frac{\phi+(1-\phi) \rho_{ux}^{(1)}}{\phi \rho_{ux}^{(1)} + (1-\phi)} \right)\left(\frac{\mu_x^{(0)}-\mu_x^{(1)}}{\sqrt{\sigma_{xx}^{(1)}}}\right) \label{eqn:ppmm2}\\
\sigma_{uu}^{(0)} &{}= 1 + \left( \frac{\phi+(1-\phi) \rho_{ux}^{(1)}}{\phi \rho_{ux}^{(1)} + (1-\phi)} \right)^2 \left( \frac{\sigma_{xx}^{(0)} - \sigma_{xx}^{(1)}}{\sigma_{xx}^{(1)}}\right). \label{eqn:ppmm3}
\end{align}
\noindent Here $\mu_x^{(j)}$ and $\sigma_{xx}^{(j)}$ are the mean and variance of the proxy $X$ for $S=j, j=\{0,1\}$ and $\rho_{ux}^{(1)}$ is the correlation between $U$ and $X$ in the respondent sample.

Insight into how the PPMM works can be seen by closer inspection of Equations~\eqref{eqn:ppmm1}-\eqref{eqn:ppmm3}. In \eqref{eqn:ppmm2}, the mean of latent $U$ for the nonresponding portion of the population $(\mu_u^{(0)})$ is the respondent mean $(\mu_u^{(1)})$, shifted by a factor that depends on the sensitivity parameter $\phi$, the strength of the proxy as captured by the correlation between $X$ and $U$ in the respondent sample $(\rho_{ux}^{(1)})$, and how different the proxy mean is for respondents $(\mu_x^{(1)})$ and nonrespondents $(\mu_x^{(0)})$. Larger differences in proxy means between respondents and nonrespondents will lead to larger shifts of the mean of $U$. The amount of shift is also governed by $\phi$, and at the two extremes of $\phi=0$ and $\phi=1$ the first term in the parentheses in \eqref{eqn:ppmm2} is $\rho_{ux}^{(1)}$ and $1/\rho_{ux}^{(1)}$, respectively. Thus, the larger the correlation $\rho_{ux}^{(1)}$, the smaller the range of the shift as $\phi$ goes from 0 to 1. If the proxy is weak, however, this term will produce a wide range for $\mu_u^{(0)}$ as $\phi$ is varied. A similar shifting occurs for the variance of $U$ for nonrespondents as seen in \eqref{eqn:ppmm3}.

For model estimation we use the Bayesian approach described by \citet{AndridgeLittle2020}, which puts non-informative priors on all identified parameters in the PPMM to obtain draws of the overall mean of $Y$ via a Gibbs sampler. Since the data contain no information to inform $\phi$, we use a Uniform(0,1) prior, which generates a 95\% credible interval for the mean of $Y$ that effectively averages over all possible values of $\phi$. The posterior median serves as an estimate of the mean of $Y$ for $\phi=0.5$, which was recommended by \citet{LittleWest2020} as a ``point index" if a single point estimate is desired under a non-ignorable response mechanism.

\section{Applying the PPMM to Estimate Vaccine Uptake}\label{sect:uptake}

As described in Section \ref{sect:ppma}, application of the PPMM requires aggregate information for covariates $Z$ that are also available in the HPS and CTIS survey data. We used the American Community Survey (ACS) 2019 data obtained via IPUMS USA \citep{IPUMSACS} for population-level data on the following covariates available in both the HPS and CTIS: age, gender, education, race, and ethnicity. The categories for all of these covariates differed slightly between HPS and CTIS, so separate estimates of the population mean and variance were made using the ACS that matched each survey; see Supplemental Table S1 for the coding of variables across data sources. We note that income was also available in both the HPS and the ACS, but as is typical for this variable it had relatively high rates of missingness in the survey data with approximately 25\% of respondents not providing their income, and thus we elected not to use this to create the proxy.

Our responding sample for each survey was taken to be the set of records that had information on vaccination status $(Y)$ and complete covariate data $(Z)$, as the PPMM requires complete data for the respondent sample. We followed the procedures used by the respective surveys when producing their vaccination estimates in terms of how missing data in $Y$ was handled. For the HPS, an individual with a missing $Y$ value was assumed to be a ``no, not vaccinated" and was included in the sample, whereas for the CTIS an individual with missing $Y$ was dropped from the sample ($\approx$6-7\%). For covariate data, the publicly available HPS data had our $Z$ variables already singly imputed (since they were part of the Census' weighting adjustments) and thus there were no records with missing $Z$ values. In contrast, the CTIS suffered from missing data for the demographic variables that came at the very end of the survey, with approximately 15\% additional records being dropped. Due to the very large size of the CTIS surveys, analysis sample sizes were still very large, ranging from 167,000 to 290,000 across weeks. We note that the survey weights provided with each survey are not used for the PPMM analyses, and instead the responding sample is treated effectively as a non-probability sample.

As previously noted, sampling fractions for both the HPS and CTIS were small and thus we used the mean and variance of $Z$ from the ACS for the nonrespondent portion of the population, though technically these values are for the full population. Additionally, we treat the means from the ACS as though they were ``known" despite them being estimates themselves; future work is needed to incorporate uncertainty about the $Z$ at the population level into PPMM estimation.

As a benchmark truth for the proportion of the population that had received at least one dose of a COVID-19 vaccine we used the vaccination uptake statistics available from the CDC as used by \citet{Bradley2021} and available via their GitHub repository (\url{https://github.com/vcbradley/ddc-vaccine-US}). As noted in \citet{Bradley2021}, this benchmark data itself is potentially subject to error, though retroactive corrections are included in these counts.

\begin{figure}[bt]
\includegraphics[width=\textwidth]{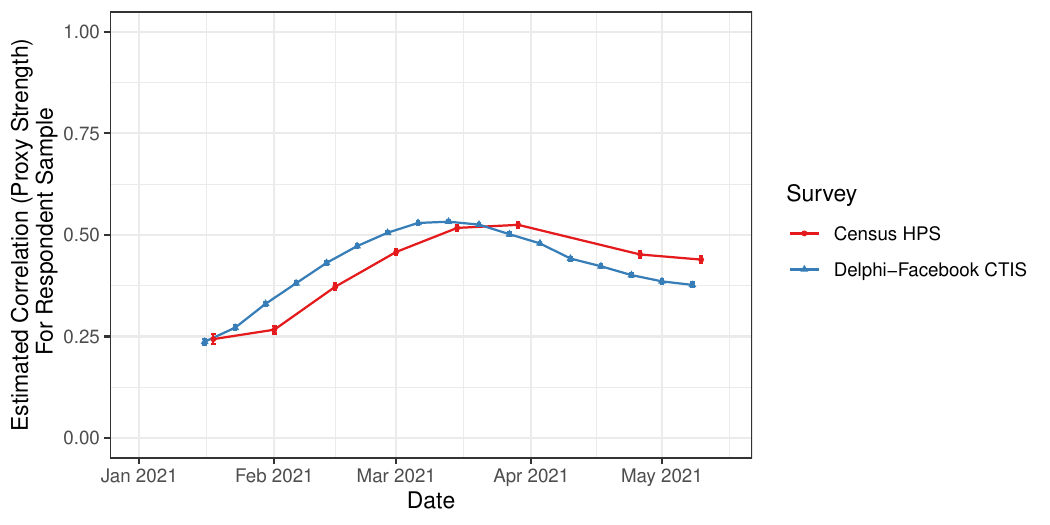}
\caption{Posterior medians for the biserial correlation $(\rho^{(1)})$ between COVID-19 vaccination uptake (binary $Y$) and proxy $X$ for the selected sample under the proxy pattern-mixture model. Bounds shown are 95\% credible intervals (too small to see for Delphi-Facebook CTIS).}
\label{fig:rhovac}
\end{figure}

Figure~\ref{fig:rhovac} shows the estimated proxy strength, i.e., the estimated correlation between $U$ and $X$ for respondents in both the HPS and CTIS during January through May of 2021. In the earlier waves, when vaccines were first available only to limited groups (e.g., older adults), the model that builds the proxy is relatively weak (around $\hat{\rho}_{ux}^{(1)}=0.25$). As vaccines became more widely available, the proxy strength increases, to a high of slightly larger than $\hat{\rho}_{ux}^{(1)}=0.5$, with a small decrease in April and May.

\begin{figure}[bt]
\includegraphics[width=\textwidth]{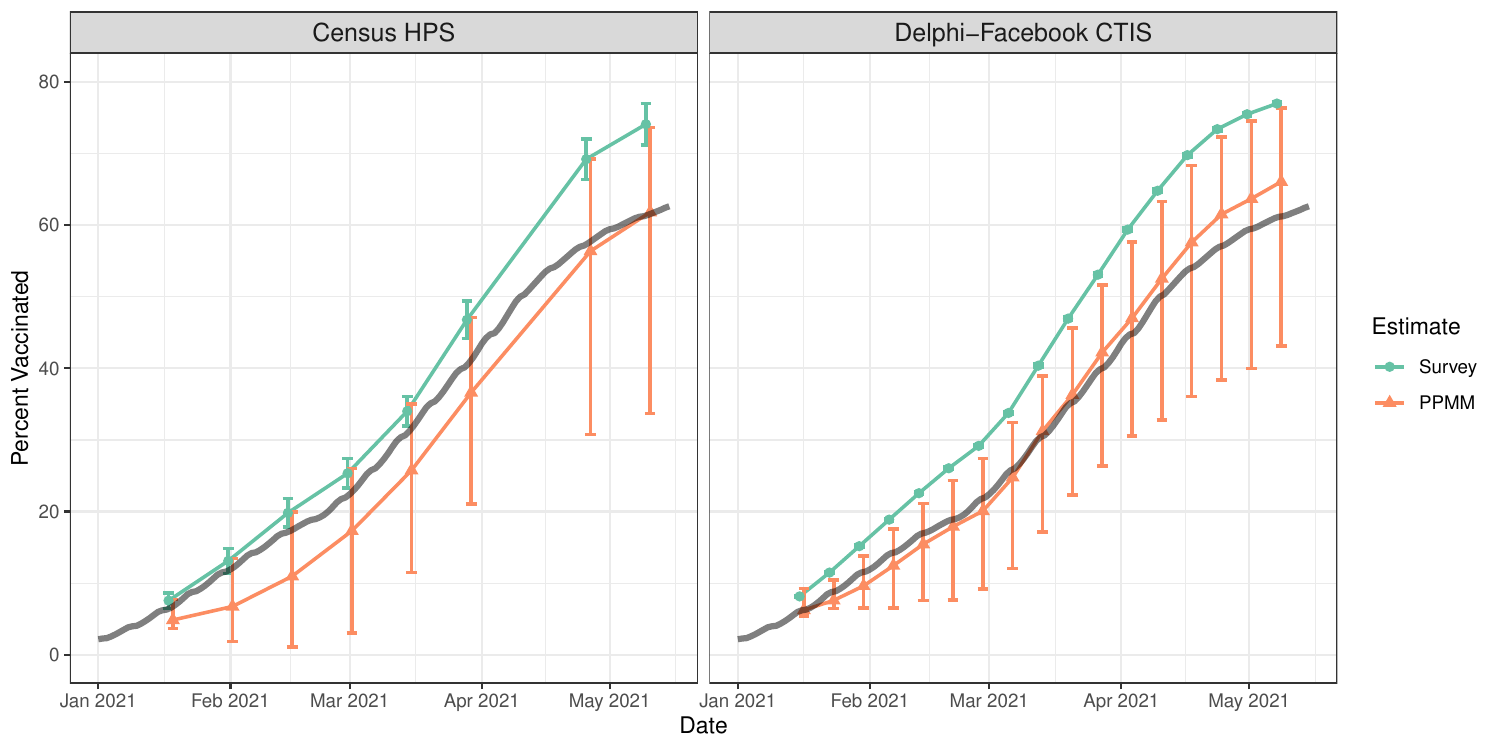}
\caption{Estimates of vaccine uptake using the proxy pattern-mixture model (PPMM) with a Uniform(0,1) prior on the sensitivity parameter $\phi$, for both the Census HPS and the Delphi-Facebook CTIS. Shown are the posterior medians with 95\% credible intervals. The grey line is the benchmark CDC data (the ``truth").}
\label{fig:vaccinated}
\end{figure}

Figure~\ref{fig:vaccinated} shows the estimates of vaccine uptake under the PPMM with a Uniform(0,1) prior on the sensitivity parameter $\phi$ for both surveys, compared to the CDC benchmark and the direct survey (weighted) estimates. Several patterns are evident in the results. First, the upper endpoint of the credible intervals corresponding to $\phi=0$ is nearly identical to the weighted estimates for HPS, which is expected since the covariates $Z$ that created the proxy are the same as those used in the weighting adjustments. For the CTIS, the interval endpoint is slightly lower than the direct estimates, as a result of our PPMM using \emph{more} information than the survey weights which only used age and gender, since education and race/ethnicity (used in the PPMM) were predictive of vaccine uptake.

Second, the PPMM credible intervals cover the benchmark truth for both surveys in all waves/weeks, while the direct survey estimates only cover the truth twice (the first two waves of the HPS). Importantly, the PPMM correctly detects the \emph{direction} of bias for both surveys in all waves/weeks, i.e., the PPMM indicates that the direct estimates were overestimating the true proportion of adults who had at least one vaccine dose. For the CTIS, the posterior median proportion (corresponding to $\phi=0.5$) is remarkably close to the truth across all waves; for the HPS this ``point index" value is too low (i.e., overcorrects the bias) in the earlier waves when the HPS direct estimates are not as biased.

Finally, the PPMM credible intervals are much wider than the confidence intervals for the survey estimates despite the very large sample sizes. This is a desirable property, since one of the problems highlighted by \citet{Bradley2021} is the ``big data paradox" of \citet[p.702]{Meng2018}: ``The bigger the data, the surer we fool ourselves." The relatively larger intervals of the PPMM reflect the strength -- or weakness -- of the proxy model. Since the covariate data $Z$ are only moderately associated with $Y$, our confidence in how much non-ignorable nonresponse bias might be present is only moderate, corresponding to larger credible intervals.

\section{Applying the PPMM to Estimate Vaccine Hesitancy}\label{sect:hesitant}

We also used the PPMM, with the same set of covariates $Z$ and same external population source, to estimate the proportion of U.S. adults who were vaccine hesitant for both the HPS and CTIS data. Individuals who reported that they would ``probably not" or ``definitely not" choose to be vaccinated or were ``unsure" (HPS only) were coded as being vaccine hesitant (see Table~\ref{tab:questions} for exact question wording and response options). Individuals who either had received a vaccine dose or who ``definitely" or ``probably" would do so were coded as not being vaccine hesitant.

Proxy strength for the models for vaccine hesitancy was relatively stable both across time and between surveys. The posterior median for $\rho^{(0)}$ for the HPS ranged from 0.392 to 0.415 across waves. For the CTIS, $\rho^{(0)}$ was largest at the earliest time point (0.391) and slightly declined across the time, with the smallest posterior median at the last time point (0.332). As such, the proxy for vaccine hesitancy was generally weaker than the proxy for vaccine uptake. The full set of estimates are available in Supplemental Figure S1.

\begin{figure}[ht]
\includegraphics[width=\textwidth]{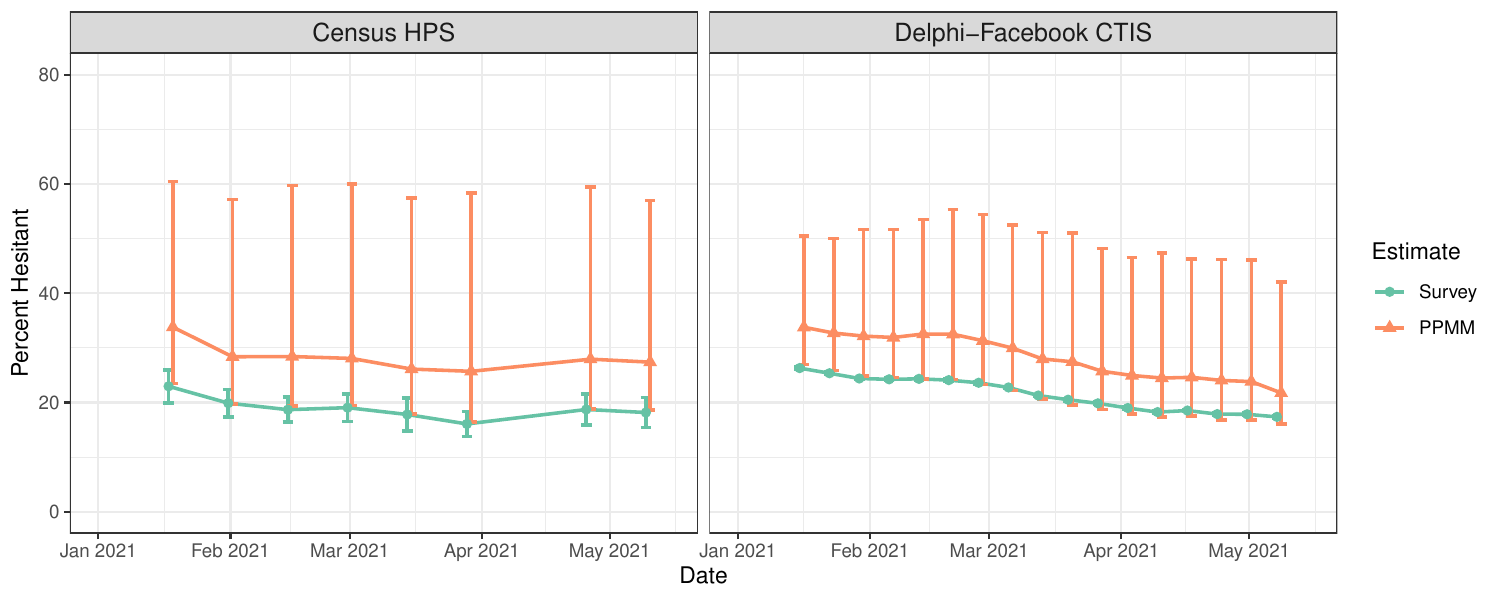}
\caption{Estimates of vaccine hesitancy using the proxy pattern-mixture model with a Uniform(0,1) prior on the sensitivity parameter $\phi$, for both the Census HPS and the Delphi-Facebook CTIS. Shown are the posterior medians with 95\% credible intervals.}
\label{fig:hesitant}
\end{figure}

Results of applying the PPMM are shown in Figure~\ref{fig:hesitant}. As one might hypothesize, given that vaccine uptake was overestimated by these surveys, the PPMM suggests that vaccine hesitancy is \emph{underestimated} by a relatively stable amount across time. Using the posterior median as a point estimate under a non-ignorable response mechanism, the results suggest that vaccine hesitancy is being underestimated by around 9 percentage points on average for the HPS and by around 7 percentage points on average for the CTIS. As expected due to the relatively weak proxy, the credible intervals are large, averaging approximately 40 percentage points wide for HPS and 30 percentage points wide for CTIS. Nonetheless, this provides some evidence that the survey estimates may be too optimistic when it comes to estimating vaccine hesitancy if nonresponse is non-ignorable.

\section{Discussion}\label{sect:discussion}

In this analysis of two large surveys that substantially overestimated vaccine uptake in the U.S. in early 2021, the PPMM correctly detected the direction of bias for all survey waves. This suggests that non-ignorable nonresponse is a plausible explanation for the bias -- individuals who were not vaccinated were less likely to respond to these surveys. In addition to correctly detecting the direction of bias, median posterior estimates from the PPMM, corresponding to $\phi=0.5$ (previously suggested as a way to obtain a single estimate under the PPMM) were remarkably accurate. For the Delphi-Facebook CTIS, PPMM estimates with $\phi=0.5$ were close to the retrospectively available benchmark truth in all survey waves. For the Census HPS, estimates for $\phi=0.5$ were very close to the truth in the last two waves, when the true bias was the largest.

The success of the PPMM in the vaccine uptake context is in part due to the fact that the factors available at the population level, i.e., demographics, were moderately predictive of the outcomes of interest. If other outcomes on the same surveys are not as strongly associated with demographic characteristics then the proxies will be weaker. Having a weak proxy means that credible intervals from the PPMM will be relatively wide, and the analysis will be less informative. Nonetheless, the present analysis highlights the fact that demographic data alone can in fact provide enough information for a meaningful sensitivity analysis and provide reasonable bounds on the potential bias.

Importantly, the data necessary for a sensitivity analysis based on the PPMM are data that would be readily available in most scenarios. The only additional data needed beyond the survey microdata itself (from respondents) are population-level means and variances for the variables that create the proxy. In most cases these would be available while the survey data is first being analyzed. In fact, in many cases these population margins will be the same as what would be used for post-survey weighting adjustments.

Another reason for the success of the PPMM in our context is that the target population is a relatively stable and clearly defined population for which summary statistics are readily available. This may not always be the case. For example, when applying the PPMM to pre-election polling we found very strong proxies $(\rho^{(0)} \ge 0.9)$  \citep{WestAndridge2023}. However, the challenge there was in defining the population of interest. A pre-election poll attempts to make inference to a dynamic population of ``likely voters." Finding aggregate data for such a population is a major challenge, unlike the relatively simple task of finding demographic summaries for all adults in the U.S. in the vaccine uptake application.

Overall, this \emph{retrospective} analysis provides evidence that the PPMM could be used as a method for \emph{prospective} assessment of the potential for non-ignorable nonresponse bias. In most cases, a benchmark truth will not be available, but this application suggests that the PPMM can in fact capture the truth in a ``real data" setting. Our analysis also provides support for \citeauthor{LittleWest2020}'s recommendation of $\phi=0.5$ as a reasonable point estimate, a ``moderately non-ignorable" mechanism that falls halfway between the ignorable $(\phi=0)$ and most extremely non-ignorable $(\phi=1)$ sensitivity bounds.

\section*{Data Availability}

Census HPS microdata are publicly available for download from \url{https://www.census.gov/data/experimental-data-products/household-pulse-survey.html}. Delphi-Facebook CTIS individual-level microdata are available to eligible academic and nonprofit researchers with fully executed data use agreements, see \url{https://dataforgood.facebook.com/dfg/docs/covid-19-trends-and-impact-survey-request-for-data-access}. The HPS data used in this paper, along with code to replicate the analyses, are available at \url{https://github.com/randridge/PPMA}, along with code only for the Delphi-Facebook analyses.

\newpage \singlespace
\bibliographystyle{rss}
\bibliography{biblio}

\newpage

\appendix
\section*{Supplemental Information}

\setcounter{table}{0}
\setcounter{figure}{0}
\renewcommand{\thetable}{S\arabic{table}}  
\renewcommand{\thefigure}{S\arabic{figure}}

\input{SUPP_INFO.tex}

\end{document}

%% file: SUPP_INFO.tex
\begin{figure}[htb]
\includegraphics[width=\textwidth]{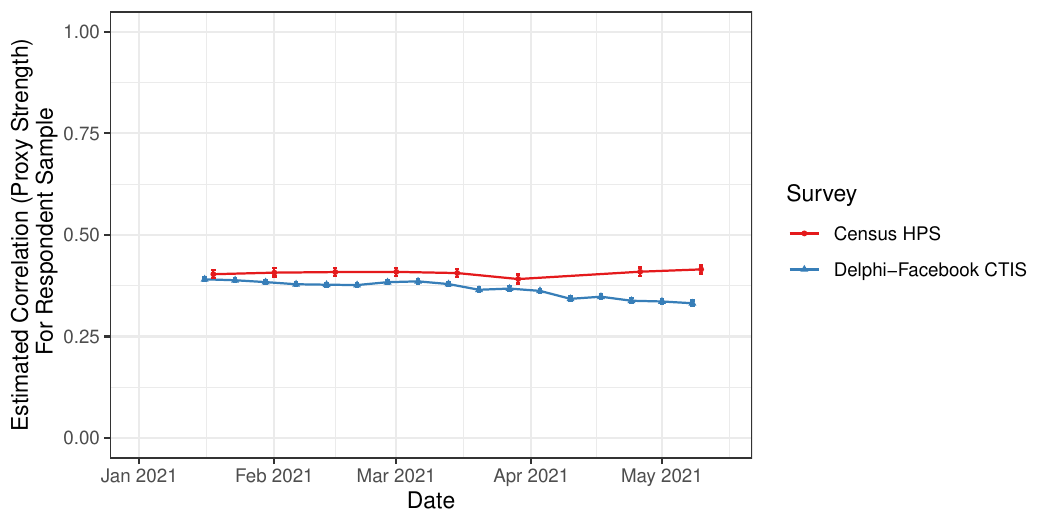}
\caption{Posterior medians for the biserial correlation $(\rho^{(1)})$ between COVID-19 vaccination hesitancy (binary $Y$) and proxy $X$ for the selected sample under the proxy pattern-mixture model. Bounds shown are 95\% credible intervals (too small to see for Delphi-Facebook CTIS).}
\end{figure}

\begin{landscape}

\begin{table}[htb]
\begin{center}
\caption{Demographic variables used to create the proxy for Census HPS and Facebook-Delphi CTIS.}
\resizebox{1.4\textwidth}{!}{%
\begin{tabular}{lll}
\toprule
\textbf{Variable} & \textbf{Census HPS} & \textbf{Delphi-Facebook CTIS} \\ \midrule
Sex & Male & Male \\
 & Female & Female OR Non-binary OR Prefer not to self-disclose OR Prefer not to answer \\ \midrule
Education & Less than high school OR Some high school OR HS grad or equiv. & Less than HS \\
 & Some college OR Associate's degree & HS grad or equivalent \\
 & Bachelor's degree & Some college OR 2 year degree \\
 & Graduate degree & 4 year degree \\
 &  & Master's degree OR Professional degree OR Doctorate \\ \midrule
Race & White alone & -- \\
 & Black alone & -- \\
 & Asian alone & -- \\
 & Any other race alone or in combination & -- \\ \midrule
Ethnicity & Not Hispanic & -- \\
 & Hispanic & -- \\ \midrule
Race/Ethnicity* & -- & Non-Hispanic White \\
 & -- & Non-Hispanic Black \\
 & -- & Non-Hispanic American Indian/Alaska Native \\
 & -- & Non-Hispanic Asian/AAPI \\
 & -- & Hispanic \\ \midrule
Age & 18-29 & 18-24 \\
 & 30-39 & 25-34 \\
 & 40-49 & 35-44 \\
 & 50-59 & 45-54 \\
 & 60-69 & 55-64 \\
 & 70+ & 65-74 \\
 &  & 75+ \\
 \bottomrule
\multicolumn{3}{l}{\small Note: Categories separated by``OR" were separate response options and are combined for analyses}\\
\multicolumn{3}{l}{\small *Race and ethnicity data were captured separated on the CTIS but were combined into a single race/ethnicity variable available with the microdata.}
\end{tabular}
}
\end{center}
\end{table}

\end{landscape}